\documentclass[twocolumn,showpacs,preprintnumbers,amsmath,amssymb,nofootinbib,floatfix,superscriptaddress]{revtex4}
\usepackage{graphicx}
\usepackage{dcolumn}
\usepackage{bm}
\begin{document}

\title{Cancelling J\" uttner Distributions for Space-like Freeze-out }
\author{K. Tamo\v si\=unas}
\email{karolis@fi.uib.no}
\affiliation{Theoretical and Computational Physics Section, Department of Physics, University of Bergen, Allegaten 55, 5007 Bergen, Norway}
\author{L. P. Csernai}
\email{csernai@fi.uib.no}
\affiliation{Theoretical and Computational Physics Section, Department of Physics, University of Bergen, Allegaten 55, 5007 Bergen, Norway}
\affiliation{MTA-KFKI, Research Inst. for Particle and Nuclear Physics, 1525 Budapest 114, P.O. Box 49, Hungary}
\date {\today}
\input{epsf}
\newcommand{\beq}{\begin{equation}}
\newcommand{\eeq}[1]{\label{#1} \end{equation}}
\newcommand{\bdm}{\begin{displaymath}}
\newcommand{\edm}{\end{displaymath}}

\begin{abstract}
We study freeze out process of particles across a three dimensional space-time
hypersurface with space-like normal. The problem of negative contribution is discussed
 with respect to
conservation laws, and a simple and practical new one-particle distribution for the
post FO side is introduced, the Cancelling J\" uttner (CJ) distribution.
\end{abstract}
\maketitle

\section{Introduction}

 Relativistic heavy ion collisions are non-equilibrium relativistic many body systems
 that can be described by different models, often based on the kinetic theory.
  The kinetic theory establishes relationship
 between macroscopic and microscopic matter properties mostly by using a one-particle distribution function.
 Kinetic theories are able to describe dilute weakly interacting systems, thus expanding
  systems where the constituent particles gradually loose contact.
 Here to simplify such a process we will assume a freeze out (FO) hypersurface in space-time,
  which can have either space- or time-like normal vector.
  We will use conservation laws of fluid dynamics to conserve energy, momentum and
 particle flow across FO hypersurface. We assume thermal equilibrium in pre FO side where the
  matter is (or can be) in quark gluon plasma (QGP) phase. For the pre FO side
   the Bag Model equation of state will be used which allows for supercooling and rapid hadronization
   afterwards.
 For the frozen out matter (post FO side) newly formed hadrons are considered, which are non interacting with each other,
 and they are described by a one-particle distribution function.
 The invariant number of conserved
 particles (world lines) crossing a surface element, $d\sigma^\mu$, of the FO hypersurface is:
   $dN=N^{\mu}d\sigma_{\mu}$.
 Thus, the total number of all particles crossing the FO hypersurface is $ N  = \int_S N^\mu d\sigma_\mu$.
 From kinetic definition of the baryon current four vector, $N^{\mu}$, we have:
 \beq
 N^\mu = \int \frac{d^3 p}{p^0}p^\mu f_{FO}(\vec{r},p;T,n,u^\mu).
 \eeq{Nmu}
 Inserting it into the equation for the total number of particles leads to the Cooper-Frye formula
 \cite{CFry}:
 \beq
 E \frac{dN}{d^3p}= \int f_{FO}(\vec{r},p;T,n,u^\mu) p^\mu d \sigma_\mu ,
 \eeq{CF}
 where $f_{FO}( \vec{r},p;T,n,u^\mu)$ is the unknown post FO phase space distribution of frozen out particles.
 The problem is to choose it's form correctly, and to determine its parameters which satisfy
 all conservation laws! This was not done correctly in the original Cooper-Frye description \cite{CFry}.
  The most used distribution  for the time-like
 normal is the J\" uttner distribution \cite{Juttner} (also called relativistic Boltzmann distribution):
 \beq
 f^{Juttner}(\vec{r}, p)=\frac{1}{(2\pi \hbar)^3}\exp{ \left( \frac{\mu(\vec{r})-p^\mu u_\mu(\vec{r})}{T(\vec{r})} \right) }
 \eeq{Jutt}
 For the time-like case $p^\mu$ and $d\sigma^\mu$ in the Cooper-Frye formula
 are both time-like vectors, thus $p^\mu d\sigma_\mu >0$, and the integrand of integral (\ref{CF}) is always positive.
 For the space-like normal vector case, $p^\mu d\sigma_\mu$ can be both positive and negative.
 This situation is a problem, because the integrand  in integral (\ref{CF}) may change sign,
 and this indicates, that part of distribution contributes to a negative current, going back
 into the front, while the other part is coming out of the front.
 The improvement of the distribution function (\ref{Jutt}) was done by introducing the cut-J\" uttner
 distribution  \cite{Anderlik}, \cite{Bugaev}, which makes $ p^\mu d\sigma_\mu$ to be non-negative in J\" uttner distribution
 by multiplying it with step function: $f^{Cut-Juttner}=\Theta(p^\mu d\sigma_\mu)f^{Juttner}$. The
  cut-J\" uttner distribution has solved the above mentioned problem formally, but the
  lack of a real
 physical solution persisted. The cut-J\" uttner distribution has an unphysical form:
 the distribution is sharply cut off which is mathematically satisfactory, but it is hardly possible
 to imagine a realistic physical process producing such a distribution. So, taking into account that
 this distribution describes physical particles  - it has to be improved. A solution to this problem
 in kinetic theory was presented in ref. \cite{KFOM}.

 In the present work we will present a new, simple distribution, called the
 Cancelling J\" uttner (CJ) distribution.
   It solves the problem of negative contributions in the Cooper-Frye
formula, and it has a smooth physically realistic form. We worked out simple relations, which
 can be used for the calculation of measurables
 and can be compared with experimental results.

\section{The Cancelling J\" uttner Distribution}

\begin{figure}[!ht]

\begin{minipage}[t]{75mm}
  \hspace{-1cm}
    \includegraphics[width=8cm]{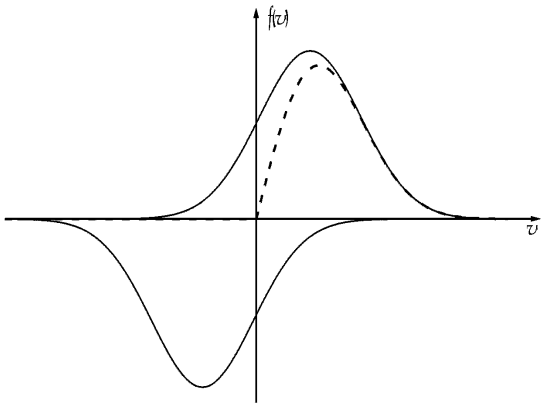}
 \caption{ Schematic view of making the Cancelling J\" uttner
 distribution: in the Rest Frame of the Front two J\" uttner
distributions (full curves) - one with negative velocity, another with positive are subtracted.
Dashed curve is the CJ distribution.\label{figure_normal} }
\end{minipage}
\end{figure}

The Cancelling J\" uttner (CJ) distribution, $f_{CJ}$, is defined by subtracting the ordinary J\" uttner
distribution (\ref{Jutt}) with negative velocity, $-v$, from original J\" uttner distribution, and
and multiplying the obtained result with the step function (figure \ref{figure_normal}):
\begin{eqnarray}
\lefteqn{f_{CJ}=\left( f^{Juttner}_R - f^{Juttner}_L \right) \Theta(p^\mu d\sigma_\mu) = {} }\nonumber\\
& & {=} \frac{\Theta(p^\mu d\sigma_\mu)}{(2\pi \hbar)^3}
\left(\exp{\frac{\mu{-}p^\mu u_\mu ^R}{T}}{-} \exp{\frac{\mu{-}p^\mu u_\mu ^L}{T}}
\right),
\label{CJ}
\end{eqnarray}
\begin{figure}[!ht]
\begin{minipage}[t]{75mm}
    \includegraphics[height=8cm, angle=270]{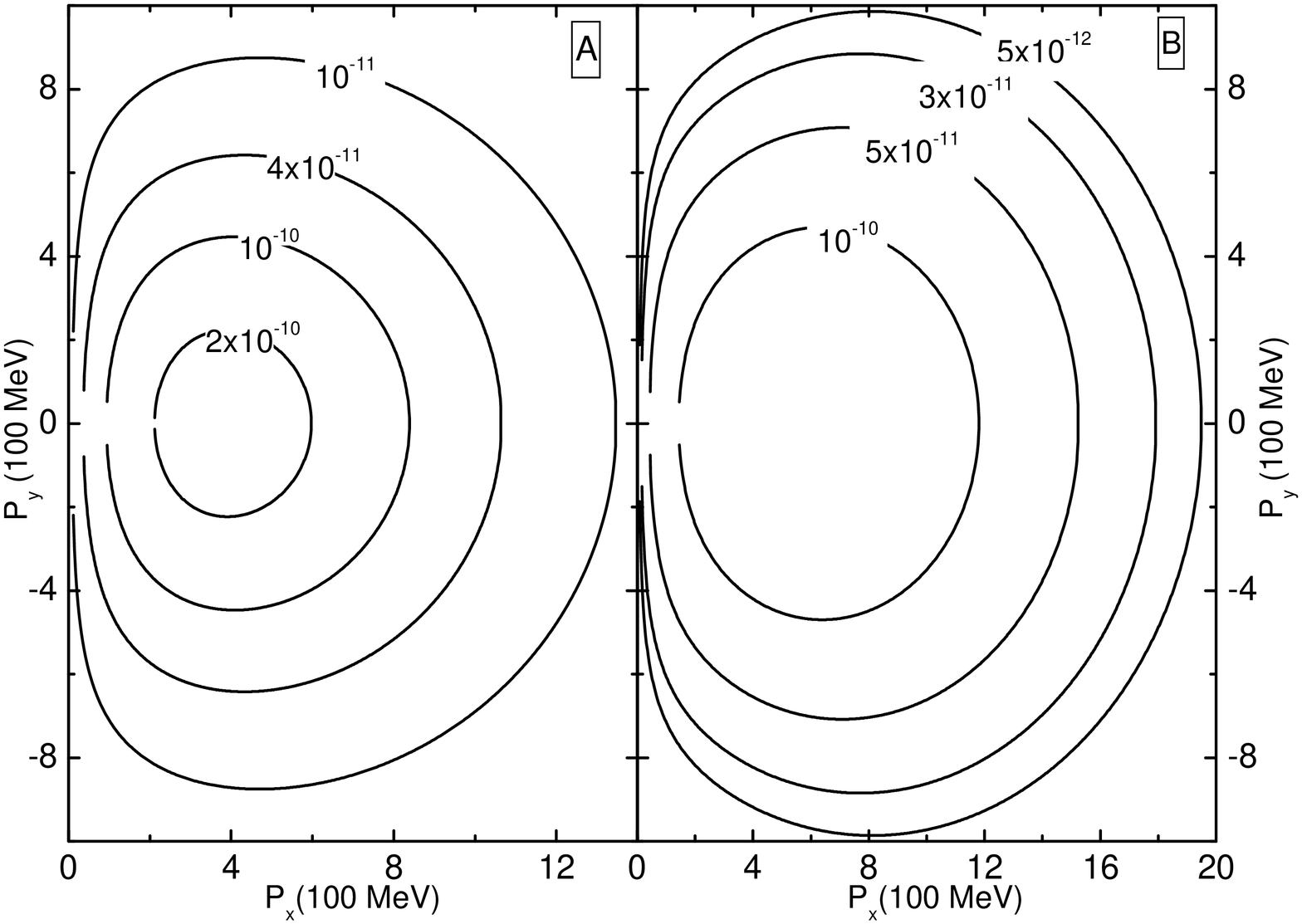}
\caption{ The post FO CJ distribution, $f_{CJ}$, in the rest frame
of the front (RFF). Baryon mass, $m=1$ GeV, temperature, $T=100$ MeV,  density, $n=0.17$ fm$^{-3}$
and velocity, $v=0.3$ c in 'A',  $v=0.5$ c in 'B'.
The values of each contour are written on the curves. The
CJ distribution resembles strongly the post FO distribution obtained in kinetic
theory, fig. 2 and 3 in ref. \cite{KFOM}.  \label{fig_CJ} }

\end{minipage}
\end{figure}
where $ u_\mu^R=(\gamma , \gamma v , 0, 0) \ \ \textrm{and} \ \ u_\mu^L=(\gamma , -\gamma v , 0, 0)$ in the rest frame of the front (RFF).
 The velocity parameter, $v$, of the CJ distribution is restricted to be positive.
When $p^\mu d\sigma_\mu = 0$ the function vanishes at the front, even without step function $\Theta$.
The role of the $\Theta(p^\mu d\sigma_\mu)$ part is just to eliminate the negative part of the distibution.

We chose this construction from two Juttner distributions
because:
 1) It automatically includes the cut, and furthermore a smooth but rapid cut, thus the unrealistic
profile of the earlier proposed cut-Juttner distribution
\cite{Bugaev} is not present.
 2) It resembles the distribution we obtained from a kinetic model quite well, in any case much better
than the cut-Juttner distribution.
 3) The formulation is still analytic and not more complicated than the one arising from the
cut-Juttner distribution.

Calculations and results will be made in the rest frame of the front (RFF),
 where the surface element is $d\sigma_\mu = (0,1,0,0)d\sigma$.
In this frame particles cannot propagate across the front in the negative $x$ direction. We assume that the pre FO side
is in thermal equilibrium. If the particle has passed the freeze out layer it cannot scatter
 back. This layer is idealized as a front or hypersurface.
In figure \ref{figure_normal}, the schematic way of making the CJ distribution is shown.
The distribution obtained this way (figure \ref{fig_CJ}) resembles strongly the one obtained numerically
in a kinetic FO model \cite{KFOM}. In the following we demonstrate the use of the Cancelling J\" uttner distribution.


\section{Conservation Laws of Fluid Dynamics}

The FO hypersurface is considered to be a discontinuity in space-time,
thus instead of derivatives of the discontinuous variables we consider the changes explicitely.
We will use the conservation laws of fluid dynamics, but instead of the differential form of
continuity equations we have to use comutator form:
\beq
[N^\mu d\sigma_\mu]=0 \  \  \textrm{and} \  \  [T^{\mu\nu}d\sigma_\mu]=0,
\eeq{conlaws}
where $[A]=A_1-A_0$, $A_1$ is a post FO quantity, and $A_0$ is the pre FO quantity. Moreover, entropy across FO hypersurface must not decrease:
\beq
[S^\mu d\sigma_\mu] \ge 0 .
\eeq{entropy}
This condition does not provide additional information about the freezing out matter,
 but it must be satisfied.

Now our task is to find out the expressions for baryon current,
energy-momentum tensor and entropy current in the post FO side.
 Having these expressions, we will insert them to
equations of conservation laws (\ref{conlaws}), and then we will have a relationship between matter properties
in pre and post FO sides. This calculation is straightforward for time-like FO where both the pre-
and post FO sides can be characterized as perfect fluids with well defined equation of state and
energy-momentum tensor (see \cite{ItoRHIC}). In the case of space-like FO with out of equilibrated post
 FO distribution this calculation has been demonstrated so far only for the
 cut-J\" uttner distribution \cite{Anderlik}.

  The post FO baryon current, $N^\nu$, can be calculated
inserting the post FO distribution (\ref{CJ}) into equation (\ref{Nmu}). The energy-momentum tensor,
$T^{\mu\nu}$, can be obtained in the same way, i.e., inserting the post FO distribution (\ref{CJ}) in to the definition:
 \beq
 T^{\mu\nu}=\int \frac{d^3p}{p^0}p^\mu p^\nu f_{FO}(\vec{r},p;T,n,u^\mu)
 \eeq{Tmunu}
 However, we do not have to calculate integral
(\ref {Nmu}) nor (\ref{Tmunu}), but instead we take the already calculated  $N^\mu$,  $T^{\mu\nu}$
 and $S^\mu$ for the post FO
 Cut-J\" uttner distribution in the reference frame of the gas (RFG), from \cite{Anderlik}:
\bdm
\begin{array}{rl}
 N^0=& \frac {\tilde n}{4} \Big[ vA+a^2j[(1+j)K_2(a)-\mathcal{K}_2(a,b)]+j\frac{b^3v^3}{3}e^{-b}\Big],\\

  N^x=& \frac{\tilde n}{8}\Big[(1-v^2)A-a^2e^{-b}\Big],\\
 T^{00}=& \frac{3\tilde n T}{2} \Big[ j\frac{a^2}{2}\Big((1+j)[K_2(a)+\frac
 {a}{3}K_1(a)]-\\
 -&\mathcal{K}_2(a,b)-\frac{a}{3}\mathcal{K}_1(a,b)\Big)+Bv \Big],\\
  T^{0x}=& T^{x0}=\frac{3\tilde n T}{4} \left[ (1-v^2)B-\frac{a^2}{6}(b+1)e^{-b} \right],\\
 T^{xx}=& \frac{\tilde n T}{2} \left[ j\frac{a^2}{2}[(1+j)K_2(a)-\mathcal{K}_2(a,b)]+v^3B \right],\\
 T^{yy}{=}& T^{zz}{=} \frac{3\tilde{n}T}4\Big[ v(1{-}\frac{v^2}3)B+{\frac{ja^2}3} \Big( (1+j)K_2(a)-\\
 -& {\cal K}_2(a,b)\Big) -{\frac{{va^2}}6}(b{+}1)e^{-b}\Big],\\

S^0= & \frac{\tilde{n}}4\Big[ (1{-}\frac \mu T)vA+6vB+(1-\frac \mu
T)a^2j\big( (1{+}j)K_2(a)-\\
- & {\cal K}_2(a,b)\big) + ja^2\big( (1{+}j)K_1(a)-{\cal K}_1(a,b)\big)\Big],
\\
S^x= & \frac{\tilde{n}}8\Big[( 1-v^2) ( 1-\frac \mu
T) A+6( 1-v^2) B-\\
-& a^2( 2+b-\frac \mu T)e^{-b}\Big],
\end{array}
\edm
 where $j=\textrm{sign}(v)$, $\tilde n=\pi T^3e^{\mu/T}(\pi\hbar)^{-3}$, $\mu$ is chemical potential
  of the pre FO matter, $ a=m/T, \ b=a/\sqrt{1-v^2}, \
 v=d\sigma_0/d \sigma_x, \  A=(2+2b+b^2)e^{-b}, \  B=(1+b+b^2/2+b^3/6)e^{-b}$ and
 \bdm
 \mathcal{K}_n(a,b)=\frac{2^n(n!)}{(2n)!}a^{-n}\int_b^\infty dx(x^2-a^2)^{n-1/2}e^{-x},
 \edm
 i.e. $K_n(a)=\mathcal{K}_n(a,a)$. All other components of $T^{\mu\nu}, N^\mu$ and $S^\mu$ vanish.

 The derivation of $N^\mu_{CJ}$ and $T^{\mu\nu}_{CJ}$ will be presented in detail in the mass
 zero limit in the next section. However, the way of deriving these quantities is the same for both cases: with
 mass zero limit, or without. The results
 of $N^\mu_{CJ}$, $T^{\mu\nu}_{CJ}$ and $S^\mu_{CJ}$ for finite mass $m$, in RFF are:
 \bdm
 \begin{array}{rl}
 N^0_{CJ}=&\gamma \frac{\tilde n}{4} \Big[3Av-va^2e^{-b}-Av^3+2a^2\Delta_2\Big],\\
 N^x_{CJ}=&\gamma \frac{\tilde n}{2} \left[a^2vK_2(a)+j\frac{b^3v^4}{3}e^{-b}\right],\\

 T^{00}_{CJ}=& 3\tilde n T\gamma^2 \Big[j\frac{a^2}{2}\left(\frac{a}{3}\Delta_1
 +(1+\frac{v^2}{3})\Delta_2\right)+\\
 +& B(\frac{v^5}{3}-4v^3+5v)-va^2(b+1)e^{-b}\Big],\\
 T^{x0}_{CJ}=& T^{0x}_{CJ}=\frac{\tilde n T\gamma^2 a^2}{2}\Big[4vK_2(a)+avK_1(a)\Big],\\
 T^{xx}_{CJ}=& 3\tilde n T \gamma^2 \Big[j\frac{a^2}{2}\left(v^2(\Delta_2+
 \frac{a}{3}\Delta_1)+\frac{1}{3}\Delta_2 \right)+\\
 +& B(-\frac{8}{3}v^3+4v)+a^2v^2(b+1)e^{-b}\Big], \\

 S^0_{CJ}=& \gamma \frac{\tilde n}{4} \Big[(3v-v^3)\left( (1-\frac{\mu}{T} )A+6B\right)+\\
+& 2a^2j\left( (1-\frac{\mu}{T} )\Delta_2+\Delta_1\right)-a^2v\left(2+b-\frac{\mu}{T}\right)e^{-b}\Big], \\
 S^x_{CJ}=& \gamma \frac{\tilde n}{2} \Big[a^2K_2(a)v(1-\frac{\mu}{T})+a^2K_1(a)v\Big],
\end{array}
 \edm
 where $\Delta_i \equiv K_i(a)-\mathcal{K}_i(a,b), \ \ i=1,2$ \footnote{when $v=0$, i.e.
  $\Delta_i=0$,
   all components of $T^{\mu\nu}$ and $N^\mu$ equal zero. }. All other components vanish.
 From these expressions, one obtains properties of the matter on the post FO side of the FO front.

 \section{Calculating the Relationship of Matter Properties}

 Now we will study in more detail the change of matter properties across the FO hypersurface with
  space-like normal, having local equilibrium in pre FO side, and non-equilibrated matter on
  the post FO side. In order to connect both sides of the FO hypersurface we will use conservation
   laws of hydrodynamics.

   We are taking baryon current and energy momentum tensor for the cut-J\" uttner distribution in the RFG
     with mass zero limit \footnote{notations of baryon current, energy-momentum tensor and entropy current with mass
    zero limit and without are the same; from now on $m=0$ is assumed.}.
     Such a approximation is possible when energies are high compared to
    the masses of particles. From \cite{Anderlik} $N^\mu_{RFG}$ and $T^{\mu\nu}_{RFG}$ read as:
    \bdm
    N^0_{RFG}=\tilde n \frac{v+1}{2}, \ \ N^x_{RFG}=\tilde n \frac{1-v^2}{4}
    \edm
 \bdm
 T^{00}_{RFG}=3\tilde nT\frac{v+1}{2}, \quad T^{0x}_{RFG}=3 \tilde nT \frac{1-v^2}{4},
 \edm
 \bdm
T^{xx}_{RFG}=\tilde nT\frac{v^3+1}{2},\quad T^{zz}_{RFG}=T^{yy}_{RFG}=\frac{T^{00}_{RFG}-T^{xx}_{RFG}}{2}.
 \edm
 To get the post FO baryon current and energy-momentum tensor for the CJ distribution in
 the rest frame of the front (RFF), we are making
 a Lorentz transformation from the RFG to the RFF of the components of the cut-J\" uttner distribution. Thus, we get $N^\mu_{RFF}$ and $T^{\mu\nu}_{RFF}$ for
 the cut-J\" uttner distribution in mass zero limit:
\bdm
N^\mu_{RFF}=\frac{\tilde n \gamma}{4} (-v^3+3v+2, v^2+2v+1, 0,  0)
\edm
\bdm
T^{00}_{RFF} = \frac{\gamma^2 \tilde n T}{4}(v^5{-}6v^3{+}2v^2{+}12v{+}6),
\edm
\bdm
T^{0x}_{RFF}=T^{x0}_{RFF}=\frac{\gamma^2 \tilde n T}{4}(-v^4{+}6v^2{+}8v{+}3)
\edm
\bdm
T^{xx}_{RFF}=\frac{\gamma^2 \tilde n T}{2}(v^3{+}3v^2{+}3v{+}1), \ \ T^{yy}_{RFF}=T^{zz}_{RFF}=T^{yy}_{RFG}
\edm
The energy-momentum tensor components $T^{yy}$ and $T^{zz}$ are not changing, because our reference frame
 is chosen in such a way, that the Lorentz transformation is influencing only time and one spatial
  component $x$. All other energy-momentum tensor components are equal to zero.

 Any of the components presented above for the CJ distribution, for
example, $N^0_{CJ}$ in the RFF frame,  can be found in such a way:
\bdm
N^0_{CJ}(v)=N^0_{RFF}(v) - N^0_{RFF}(-v).
\edm
This follows from the definition of the CJ distribution function, eq. (\ref{CJ}).
Thus, the components of the baryon current of the CJ distribution take the form:
\beq
N^\mu_{CJ}=\tilde n \gamma \left( \frac{-v^3 + 3v}{2}, v, 0, 0 \right),
\eeq{Nmucj}
and the components of the energy-momentum tensor:
\beq
T^{00}_{CJ}=\tilde n T\gamma^2(v^5{-}3v^3{+}6v), \ \
T^{0x}_{CJ}=T^{x0}_{CJ}=4\tilde n T \gamma^2 v,
\eeq{T00cj}
\beq
T^{xx}_{CJ}=\tilde n T \gamma^2(v^3{+}3v), \ \ T^{yy}_{CJ}=T^{zz}_{CJ}=\tilde n T\frac{{-}v^3+3v}{2}.
\eeq{Txxcj}
Now we can introduce flow, particle density, energy density and entropy density for the frozen-out matter.
 The four-flow of the particles, $u_\mu$, can be described in different ways:
 i) Eckart's definition:
 \beq
 u_{flow,E}^\mu=\frac{N^\mu_{CJ}}{\sqrt{N^\nu_{CJ} N_{\nu, CJ}}},
 \eeq{ueckart}
  where the flow $u_\mu^E$ is tied to conserved particles.
 ii) Landau's definition:
 \beq
 u_{flow,L}^\mu=\frac{T_{CJ}^{\mu\nu}u_\nu}{\sqrt{u_\rho T^{\rho\nu}_{CJ}u_\nu}},
 \eeq{ulandau}
 where the flow is tied to energy flow. In case of Cancelling J\" uttner distribution
 there two definitions do not give an identical result, because the CJ distribution is not
 an equilibrium distribution. In further calculation let us use Eckart's definition of the
 flow.
 Having expression for the flow, we can obtain other macroscopic quantities.
  Invariant scalar density $n$ is:
  \beq
 n=N^\mu_{CJ} u_\mu^E.
  \eeq{dens}
Energy density, $e$, and entropy density, $s$, reads as:
\bdm
e^E=u_\mu^E T^{\mu\nu} u_\nu^E, \ \ s^E=S^\mu u_\mu^E.
\edm
\begin{figure}[!ht]
\begin{minipage}[t]{75mm}
  \hspace{-0.6cm}
    \includegraphics[height=8cm, angle=270]{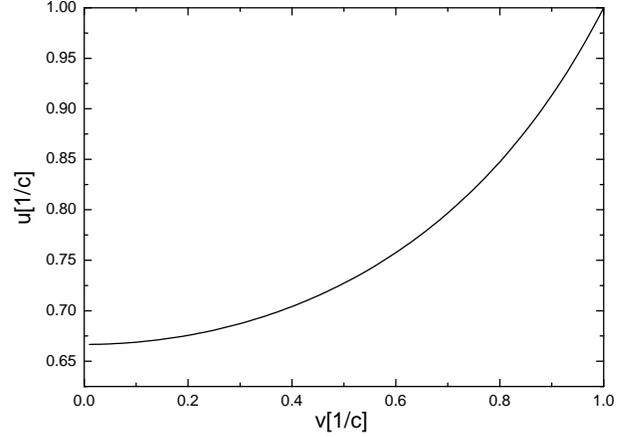}
 \caption{The Eckart's flow velocity, $u$, of the frozen-out particles, described by CJ distribution in RFF, versus velocity parameter of the distribution, $v$.  \label{flow-pr} }
\end{minipage}
\end{figure}

\begin{figure}[!ht]
\begin{minipage}[t]{75mm}
  \hspace{-0.6cm}
    \includegraphics[height=8cm, angle=270]{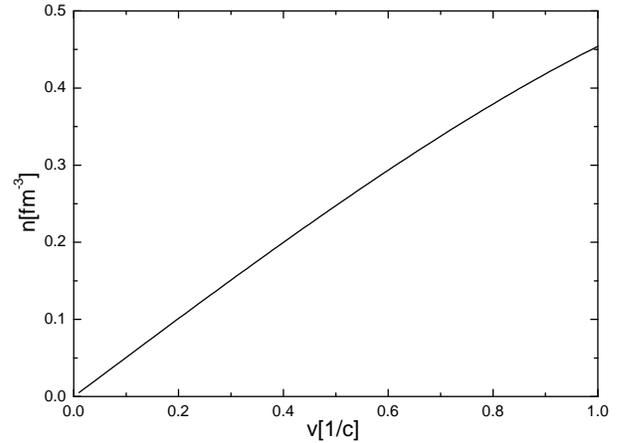}
 \caption{The proper density of the frozen-out particles, $n$, described by CJ distribution, versus
 the velocity parameter of the distribution, $v$, with fixed chemical potential, $\mu=10MeV$, and temperature \em{parameter}, $T=100MeV$. \label{dens-pr} }
\end{minipage}
\end{figure}
As we can see in figure \ref{flow-pr}, where the flow velocity is plotted versus velocity parameter of
the CJ distribution, the Eckart's flow has the minimum value $0.66 c$ in the mass zero limit.
Figure \ref{dens-pr} shows the post FO density ,$n^E$, dependece on velocity parameter, $v$.

\section{Freeze-out From QGP}

We must find expressions for the $N^\mu_0 $ and $T^{\mu\nu}_0$ in the pre FO side to use
conservation laws (\ref{conlaws}) across the surface element $d\sigma_\mu$. For perfect fluids the
energy-momentum tensor (in any reference frame) can be written as \cite{ItoRHIC}:
\beq
T^{\mu\nu}=(e+p)u^\mu u^\nu -pg^{\mu\nu},
\eeq{T_0}
where $e$ is the energy density and $p$ is pressure in the local rest (LR) frame.
For the baryon current, we use Eckart's definition for flow:
 $u^\mu=N^\mu_{(0)}/\sqrt{N^\nu_{(0)} N_\nu^{(0)}}$.
From this definition follows:
\beq
N_{(0)}^\mu=n_0 u^\mu_{(0)},
\eeq{N_0}
where $u^\mu_{(0)}=(1,0,0,0)$ and the invariant scalar, $n_0=\sqrt{N^\mu_{(0)} N_\mu^{(0)}}$,
 is the particle density.

To determine the parameters of the CJ distribution we will
use the conservation laws introduced by eq. (\ref{conlaws}). By using the
conservation law for the four-current $[N^\mu \ d\sigma_\mu]=
N^\mu_{CJ}d\sigma_\mu - N^\mu_{(0)}d\sigma_\mu^{(0)}=0$, where $d\sigma_\mu^{(0)}=\gamma_0(v_0,1,0,0)$ in the pre FO local rest frame, and
$d\sigma_\mu=(0,1,0,0)$ in the RFF, we get the equation:
\beq
\tilde n \gamma v - n_0 \gamma_0 v_0 = 0.
\eeq{1eq}
 Note that post and pre FO expressions have to be in the same reference frame, in order to use conservation
 laws correctly. Nevertheless, $N^\mu_0$ and $d\sigma_\mu^{(0)}$ do not have to be transformed to the RFF,
 because the baryon number crossing the surface is an invariant scalar.
 However, the energy-momentum current, $T^{\mu\nu}d\sigma_\mu$,  is not an
 invariant scalar, thus it must be Lorentz transformed to the RFF. After the transformation $T^{\mu\nu}$ gets a form:
 \bdm
  T^{\mu\nu}_{(0)RFF}{=}\left( \begin{array}{cccc} \gamma^2_0(e_0{+}v^2_0p_0) &
\gamma^2_0 ({-}e_0v_0{-}p_0v_0)&0 &0 \\ \gamma^2_0 ({-}e_0v_0{-}p_0v_0) &
\gamma^2_0(e_0v_0^2{+}p_0) & 0 & 0  \\ 0 & 0 & p_0 & 0 \\
0 & 0 & 0& p_0 \\
\end{array} \right).
\edm
 Using the conservation laws for the energy-momentum tensor, $[T^{\mu\nu} \
d\sigma_\mu]=T^{\mu\nu}_{CJ}d\sigma_\mu - T^{\mu\nu}_{(0)}d\sigma_\mu =0$, we obtain two more equations:
\beq
4 \gamma^2 \tilde nTv - \gamma_0^2\left(v_0e_0+v_0p_0
\right) =0
\eeq{2eq}
\beq
\gamma^2 \tilde n T(v^3+3v)-\gamma_0^2
\left( v_0^2e_0+p_0 \right)=0.
\eeq{3eq}
Entropy should not decrease, $[S^\mu d\sigma_\mu]=S^\mu_{CJ}d\sigma_\mu - S^\mu_{(0)}d\sigma_\mu^{(0)} \geq 0$, it leads to the condition:
\beq
\tilde n \gamma v \left(1- \frac{\mu}{T} \right)- s_0 \gamma_0 v_0 \ge 0,
\eeq{4eq}
which must be satisfied.

The Bag Model Equation of State (EoS) for quark gluon plasma is used to describe the pre FO matter state.
It is assumed that quarks and gluons exist in perturbative vacuum, where plasma contains
$2 (N_c^2-1)$ gluons and $2N_c N_f$ quarks ($N_c$ and $N_f$ are the number of colors and number of flavors) \cite{Sh80}.
 The Bag Model EoS is based on Stefan-Boltzmann EoS, including a bag constant $B$.
  From \cite{ItoRHIC} pre FO quantities are expressed as:
\bdm
e_0=\left(\frac{37}{30}\pi^2T_0^4+\frac{1}{3}\mu^2T_0^2+\frac{1}{54\pi^2}\mu^4+\Lambda_B^4   \right)\frac{1}{(\hbar c)^3}
\edm
\bdm
n_0=\frac{2}{9}\left(\mu^2T_0^2+\frac{1}{9\pi^2}\mu^3\right)\frac{1}{(\hbar c)^3}
\edm
\bdm
s_0=\left(\frac{74}{45}\pi^2T_0^3+\frac{2}{9}\mu^2T_0^2 \right)\frac{1}{(\hbar c)^3}
\edm
\bdm
p_0=\frac{e_0}{3}-\frac{4}{3}B, \quad \textrm{where} \quad
B=\frac{\Lambda_B^4}{(\hbar c)^3}.
\edm
Here $\mu$ is baryon chemical potential.

Using above and equations (\ref{1eq} - \ref{3eq}) we can evaluate how post FO
 matter properties depend on the pre FO side matter properties.
 It is enough to fix four quantities in the Bag Model EoS to obtain the properties of the post FO matter.
 To describe the pre FO side we will use:  initial temperature, $T_0$, bag constant, $\Lambda_B$,
   initial baryon density, $n_0$, and initial velocity, $v_0$, to have
  pre FO energy density, pressure, entropy density and baryon chemical potential.

   Nevertheless, the CJ distribution has a negative aspect: not for all initial values it is possible to calculate post FO matter parameters. There are two reasons:
    1. Entropy must not decrease, 2. The maximum of the J\" uttner distribution function must be on
     the positive velocity side. When maximum of the J\" uttner
 distribution is at velocity zero ($f_{Juttner}(v=0)|_{RFF}=\max{f_{Juttner}(v)|_{RFF}}$),
  the CJ distribution is a 'zero' function ($f_{CF}(v){=}0, \forall \  v$).
 This problem depends on how pre FO properties such as velocity, $v_0$, density,
  $n_0$, bag constant, $\Lambda_B$, and initial temperature, $T_0$, are chosen.

\begin{figure}[!ht]
\begin{minipage}[t]{75mm}
  \hspace{-0.6cm}
    \includegraphics[height=8cm, angle=270]{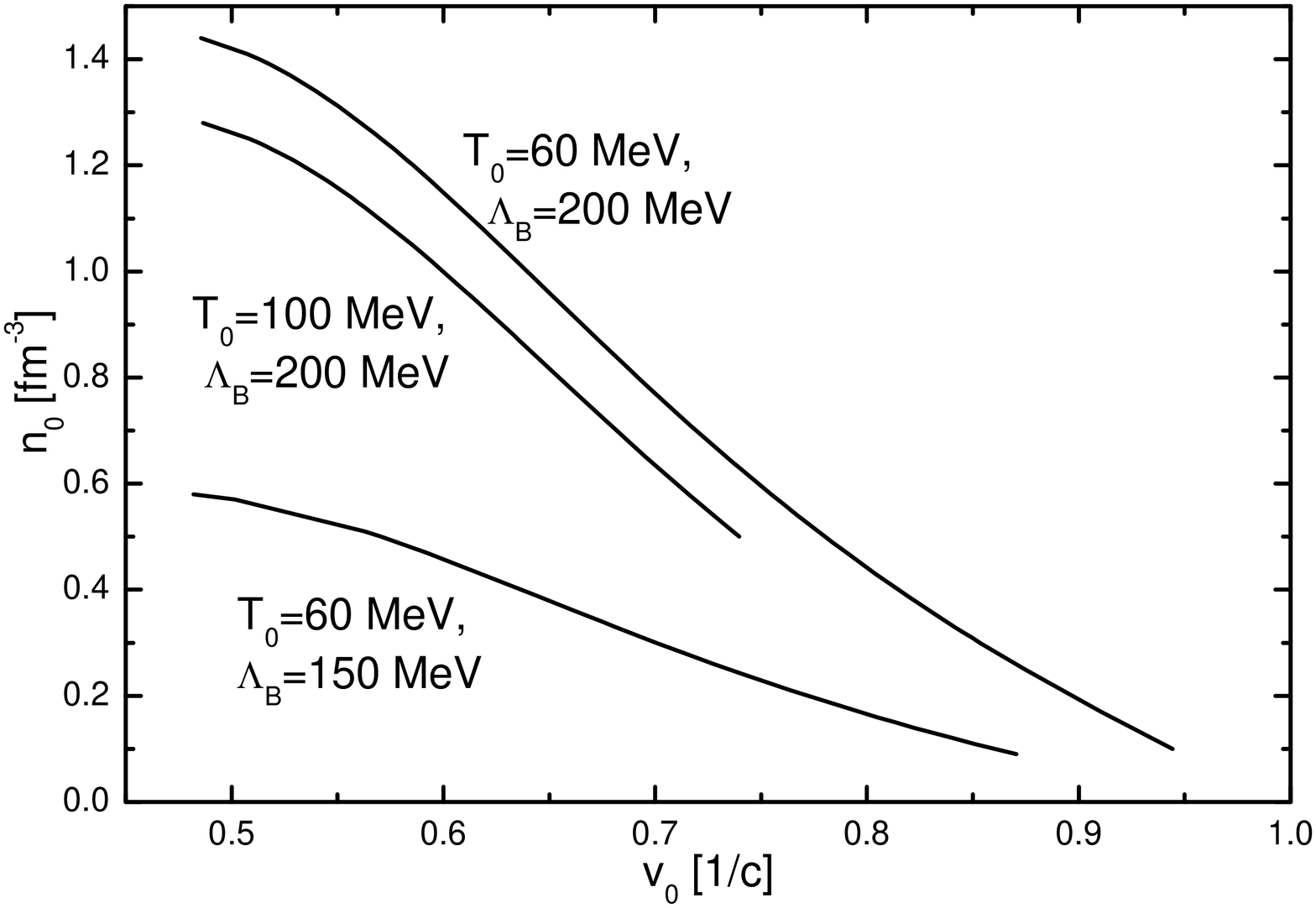}
 \caption{The boundaries of the pre FO density, $n_0$, and velocity,
$v_0$, to have correct post FO distribution function, $f_{CJ}$, in the rest frame of the front (RFF)
 with fixed initial temperature, $T_0$, and bag constant, $\Lambda_B$.
 The CJ distribution is applicable for initial conditions which are to the right from the curves,
  i.e. for large flow velocities in RFF. \label{bound} }
\end{minipage}
\end{figure}
  The boundary conditions for pre FO velocity, $v_0$ and density,
  $n_0$, with fixed bag constant, $\Lambda_B$, and initial temperature, $T_0$, are shown in figure
   \ref{bound}. Such a conditions for the pre FO matter, which are to the right
   from the curves in figure \ref{bound}, have to be satisfied for the CJ distribution because we
    are dealing with a FO hypersurface with space-like normal. The shape of the curves are influenced
  by the way of making the CJ distribution i.e. when the post FO velocity becomes imaginary.
   The cutoff of the curves is influenced by the entropy condition. In order to have perspicuity
    we compare two curves: one with initial temperature $T_0=60$ MeV, another with $T_0=100$ MeV
 ($\Lambda_B=200$ MeV for both curves). Having higher temperature, the density (or velocity) can be
 lower than in the lower temperature case. On the other hand, to insure the entropy condition, the
 density in the higher temperature case must be higher (more than $0.5$ fm$^{-3}$) than
  in the case of lower temperature, where it must be just more than $0.1$ fm$^{-3}$.
   Temperature is influencing
 the length of the curve and moves it to the left when temperature increases, and bag constant is
 changing the gradient of the curve in the $v_0, n_0$ plane.

  Now, having boundary conditions we can start calculating matter properties of the post FO side.
  From equations (\ref{1eq} - \ref{3eq}) we get one equation for the post FO velocity parameter:
\bdm
  v=\sqrt{\frac{4(v_0^2 e_0+p_0)}{v_0(e_0+p_0)}-3}.
\edm
Note, that post FO velocity, $v$, temperature, $T$, and density, $\tilde n$, are not physical
quantities - they are parameters of the CJ distribution.
\begin{figure}[!ht]
\begin{minipage}[t]{75mm}
  \hspace{-1cm}
    \includegraphics[height=8cm, angle=270]{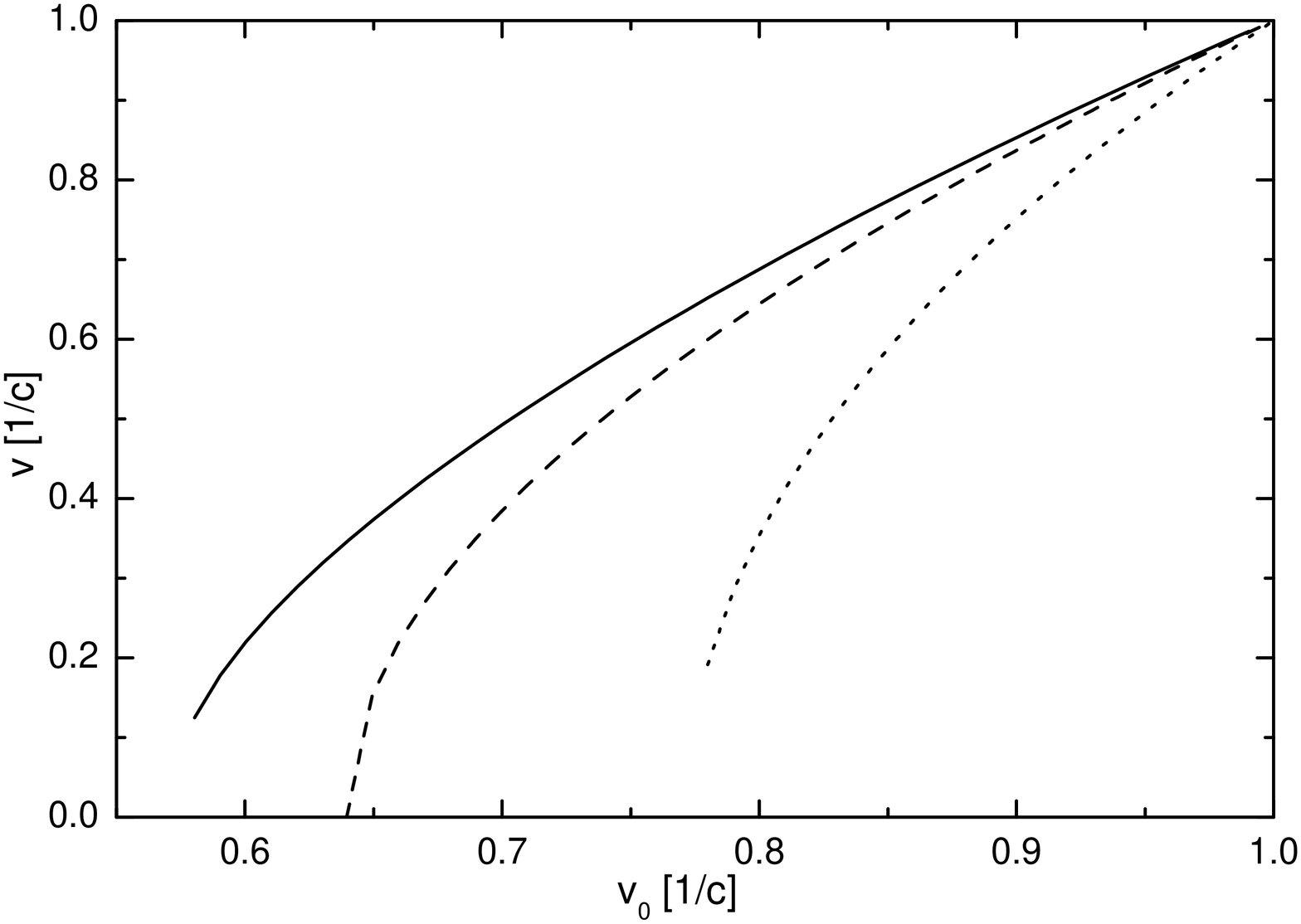}
 \caption{ Change of the velocity parameter
across the FO hypersurface, in RFF, described by the CJ distribution for the different initial
 parameters: 1. $n_0=0.5$ fm$^{-3}$, $\Lambda_0=150$ MeV full curve;
  2. $n_0=1$ fm$^{-3}$, $\Lambda_0=200$ MeV dashed curve;
  3. $n_0=0.2$ fm$^{-3}$, $\Lambda_0=150$ MeV dotted curve and $T_0=60$ MeV for all curves.\label{figure_velo}}
\end{minipage}
\end{figure}
As we see in figure \ref{figure_velo} the post FO velocity parameter is all the time smaller than the pre FO
velocity.

\begin{figure}[!ht]
\begin{minipage}[t]{75mm}
  \hspace{-1cm}
  \includegraphics[height=8cm, angle=270]{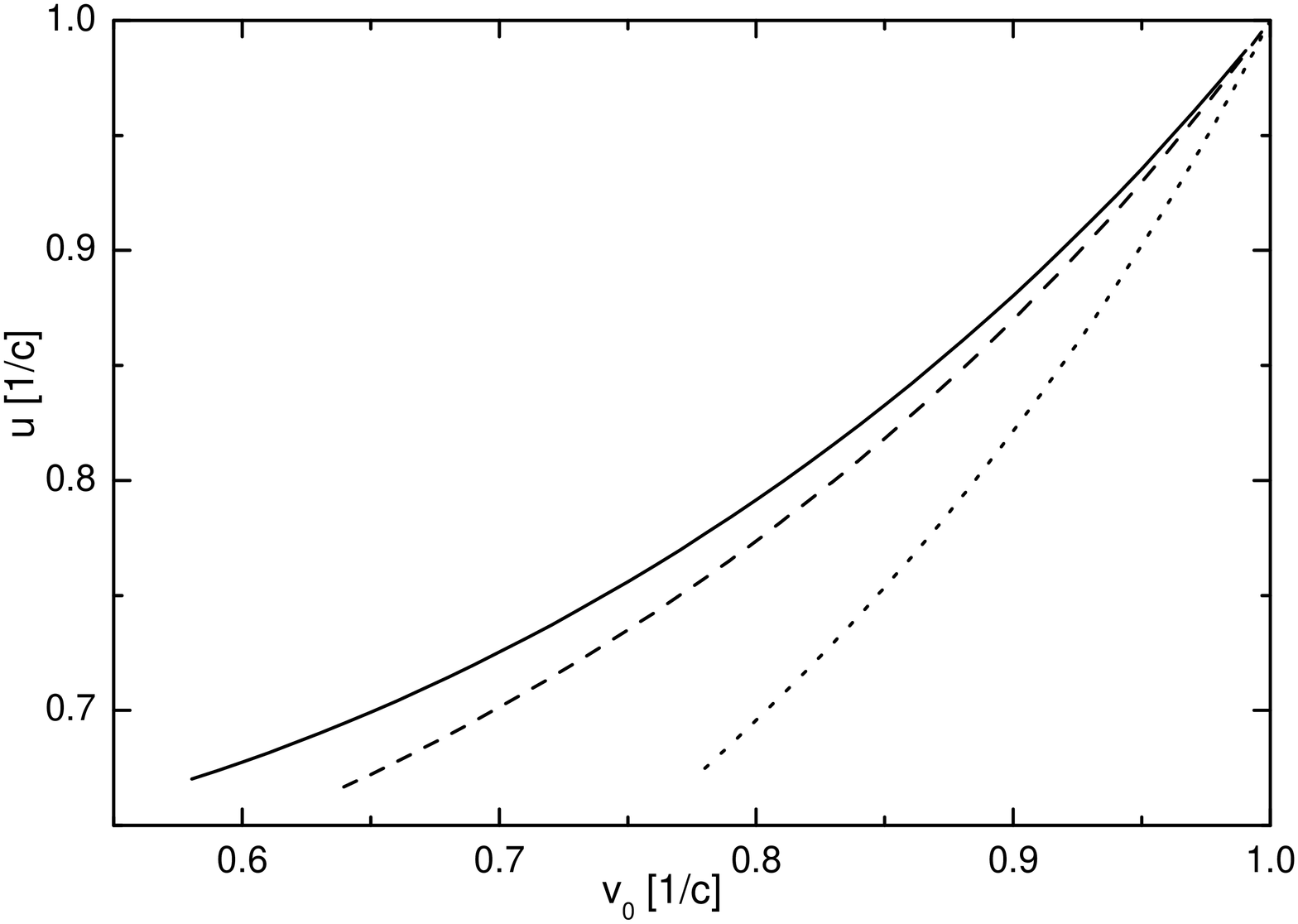}
 \caption{ The dependence of the post FO baryon flow on the pre FO
velocity for the different initial parameters:
 1. $n_0=0.5$ fm$^{-3}$, $\Lambda_0=150$ MeV full curve;
  2. $n_0=1$ fm$^{-3}$, $\Lambda_0=200$ MeV dashed curve;
  3. $n_0=0.2$ fm$^{-3}$, $\Lambda_0=150$ MeV  dotted curve and $T_0=60$ MeV for all curves. \label {fig_flow}}
\end{minipage}
\end{figure}
The post FO flow velocity, $u_{flow}$, is calculated using Eckart's
definition of the flow using formula (\ref{ueckart}).
Figure \ref{fig_flow} shows the post FO baryon flow velocity dependence on the pre FO velocity,
 $v_0$.
We observe, that the flow does not decrease in the same way as the post FO velocity parameter, $v$.
 Only in the case of low initial density ($n_0=0.2$ fm$^{-3}$ in figure \ref{fig_flow}) we have
 decrease of velocity going from pre to post FO side.
To calculate the final baryon density we are using the equation (\ref{dens}),
 which can be rewritten for the case of Eckart's flow:
\bdm n=\sqrt{N^\mu_{CJ} N_{\mu, CJ}}.   \edm
  \begin{figure}[!ht]
\begin{minipage}[t]{75mm}
  \hspace{-1cm}
    \includegraphics[height=8cm, angle=270]{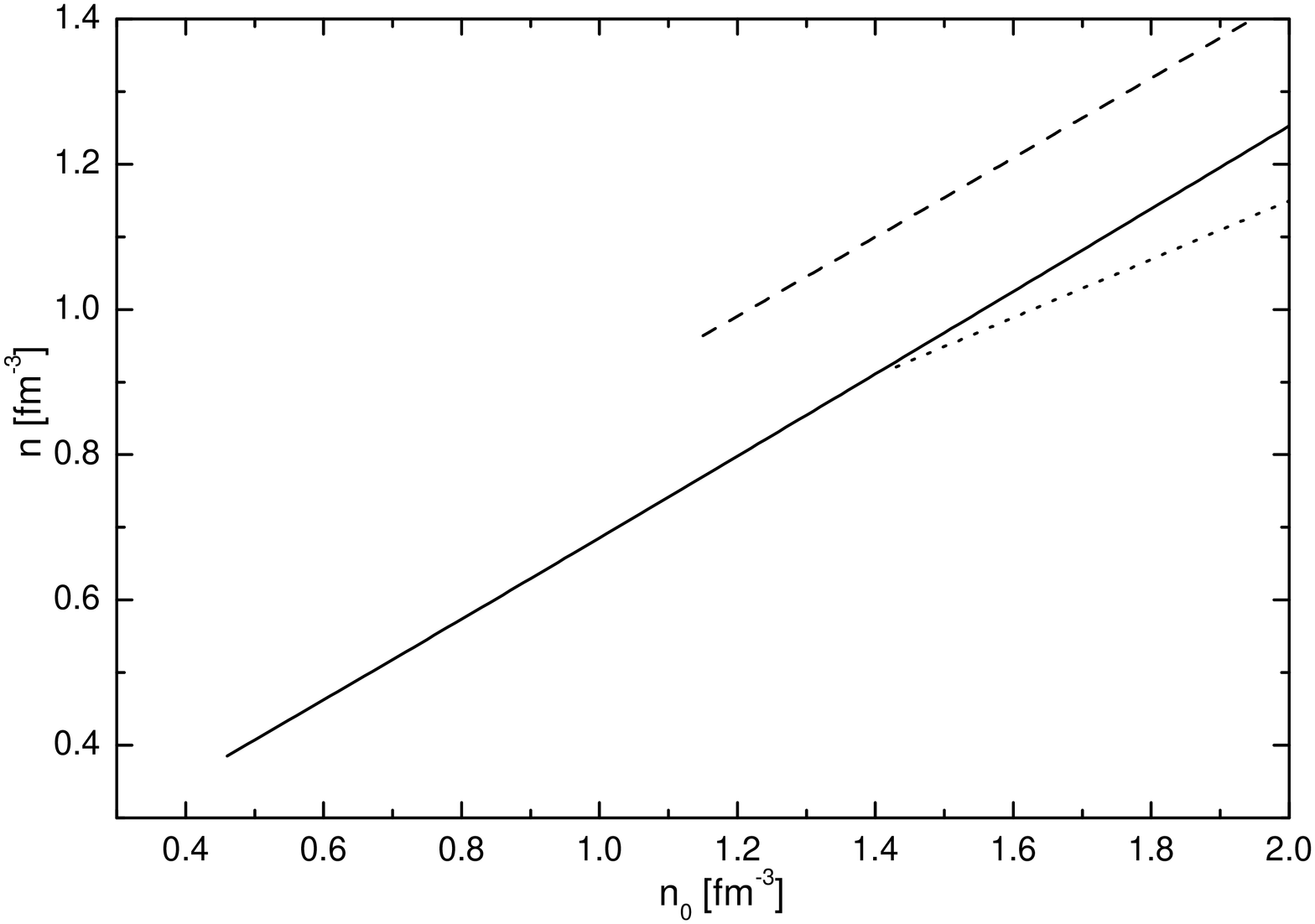}
 \caption{ The change of the baryon density across the FO hypersurface for
different initial parameters:
 1. $v_0=0.6$ c, $\Lambda_0=150$ MeV full curve;
  2. $v_0=0.6$ c, $\Lambda_0=200$ MeV dashed curve;
  3. $v_0=0.5$ c, $\Lambda_0=200$ MeV dotted curve and $T_0=60$ MeV for all curves. \label {fig_dens}}
\end{minipage}
\end{figure}
From the results presented in the figure \ref{fig_dens} it is seen
 that the baryon charge density in the post FO side is
decreasing compared to the density in the pre FO side. The two lines are parallel to each other
if initial velocity, $v_0$, and initial temperature, $T_0$, are the same.

 \section{Conclusions}

We have studied the problem of negative contribution in the Cooper-Frye (\ref{CF}) formula for the
3-dimentional hypersurface with space-like normal. The Cancelling J\" uttner distribution function was
 suggested as a solution for this problem. We have showed the applicability and properties of
  the CJ distribution by using the Bag Model equation of state for QGP and
   conservation laws of hydrodynamics
  across the FO hypersurface.

  The CJ distribution solves the problem of negative contribution in the Cooper-Frye
  formula  and has a physical form unlike the cut-J\" uttner distribution.
   In future studies one should analyze, if the conditions of applicability of CJ distribution are satisfied
   in space-like hypersurfaces obtained from hydrodynamical calculations.

   It will be interesting to see how large corrections this procedure will give to the results
   obtained from simple Cooper-Frye description.

   \begin{acknowledgments} This research has been supported by a Marie Curie Fellowship of the European
   Community programme "ECT* Doctoral Training Programme in Nuclear Theory and
Related Fields" under contact number HPMT-CT-2001-00370.
\end{acknowledgments}

\end{document}